**Packaging, containerization, and virtualization of computational omics methods: Advances, challenges, and opportunities**

Mohammed Alser[1], Sharon Waymost[2], Ram Ayyala[3,4], Brendan Lawlor[5], Richard J. Abdill[6], Neha Rajkumar[7], Nathan LaPierre[2], Jaqueline Brito[4], André M. Ribeiro-dos-Santos[8], Can Firtina[1], Nour Almadhoun[1], Varuni Sarwal[2], Eleazar Eskin[2,9,10], Qiyang Hu[11], Derek Strong[12], Byoung-Do (BD) Kim[12], Malak S. Abedalthagafi[13,14,15*], Onur Mutlu[1,*], Serghei Mangul[4,*]

[1]Department of Information Technology and Electrical Engineering, ETH Zürich, 8092 Zürich, Switzerland
[2]Department of Computer Science, UCLA, Los Angeles, CA 90095, USA
[3]Department of Translational Biomedical Informatics, University of Southern California, Los Angeles, CA 90089, USA
[4]Department of Clinical Pharmacy, University of Southern California, Los Angeles, CA 90089, USA
[5]Department of Computer Science, Munster Technological University, Cork, T12 P928, Ireland
[6]Department of Genetics, Cell Biology, and Development, University of Minnesota, MInneapolis, MN 55417
[7]Department of Bioengineering, UCLA, Los Angeles, CA 90095, USA
[8]Institute for Systems Genetics, NYU Grossman School of Medicine, New York, New York 10016, USA
[9]Department of Computational Medicine, University of California, Los Angeles, CA 90095, USA
[10]Department of Human Genetics, University of California, Los Angeles, CA 90095, USA
[11]Office of Advanced Research Computing, UCLA, Los Angeles, CA 90095, USA
[12]Center for Advanced Research Computing, University of Southern California, Los Angeles, CA 90089, USA
[13]Genomics Research Department, Saudi Human Genome Project, King Fahad Medical City and King Abdulaziz City for Science and Technology, Riyadh, Saudi Arabia
[14]King Salman Center for Disability Research, Riyadh 12512, Saudi Arabia
[15]College of Medicine, Imam Mohammad Ibn Saud Islamic University (IMSIU), Riyadh, Saudi Arabia

*These authors jointly supervised this work

**Abstract**
Omics software tools have reshaped the landscape of modern biology and become an essential component of biomedical research. The increasing dependence of biomedical scientists on these powerful tools creates a need for easier installation and greater usability. Packaging, virtualization, and containerization are different approaches to satisfy this need by wrapping omics tools in additional software that makes the omics tools easier to install and use. Here, we systematically review practices across prominent packaging, virtualization, and containerization platforms. We outline the challenges, advantages, and limitations of each approach and some of the most widely used platforms from the perspectives of users, software developers, and system administrators. We also propose principles to make packaging, virtualization, and containerization of omics software more sustainable and robust to increase the reproducibility of biomedical and life science research.

**Introduction**
High-throughput assays and subsequent computational omics analyses have led to an ever-growing, potentially overwhelming amount of biological data. Common omics approaches include genomics, proteomics, metabolomics, and transcriptomics, which are used to study genomes, proteins, metabolites, and gene transcription, respectively. Bioinformatics community has developed a multitude of software tools to leverage increasingly large and complex omics datasets[1–8]. These tools have reshaped the landscape of modern biology and become an essential component of biomedical research[9–12]. To use omics software tools, users usually need to first locate the desired tool and download it from an online repository (**Figure 1**). Then, the tool must be installed on a host computer that will run the desired analysis. The user must determine any potential dependencies and install them one after another on the host computer[13]. Dependencies are additional pieces of software that must be installed prior to the omics software being installed. The use of dependencies can ease software development and make software tools more effective, as dependencies are usually well tested, efficient, and optimized for performance[14].

The ability of many independent researchers to install and successfully run omics software tools supports scientific integrity by ensuring transparency and reproducibility of research results[15–18]. Reproducibility in this context refers to the ability to reproduce published findings by running the same omics tool with the same data used in a published study. To ensure reproducibility, researchers are encouraged, and often required, to share both raw data and installable omics software (and usually its source code) on open-access repositories such as GitHub, Dryad, or Zenodo. Source code is a collection of pieces of software written by developers using a human-readable programming language. Users intending to use omics software need to follow compilation (or building) steps, usually provided in an installation guide, for turning source code of omics software into machine code (called binary file or executable) that a computer can understand and execute. However, omics software tools often fail when reproduction is attempted because of the many challenges associated with their installation and use[19,20]. Some such common challenges include the absence of standardized interfaces, installation procedures, and data formats; a user's limited computational skills;

missing dependencies; different prerequisites and restrictions for various pieces of software; operating system (OS) incompatibility; and lack of user permissions on the host computer.

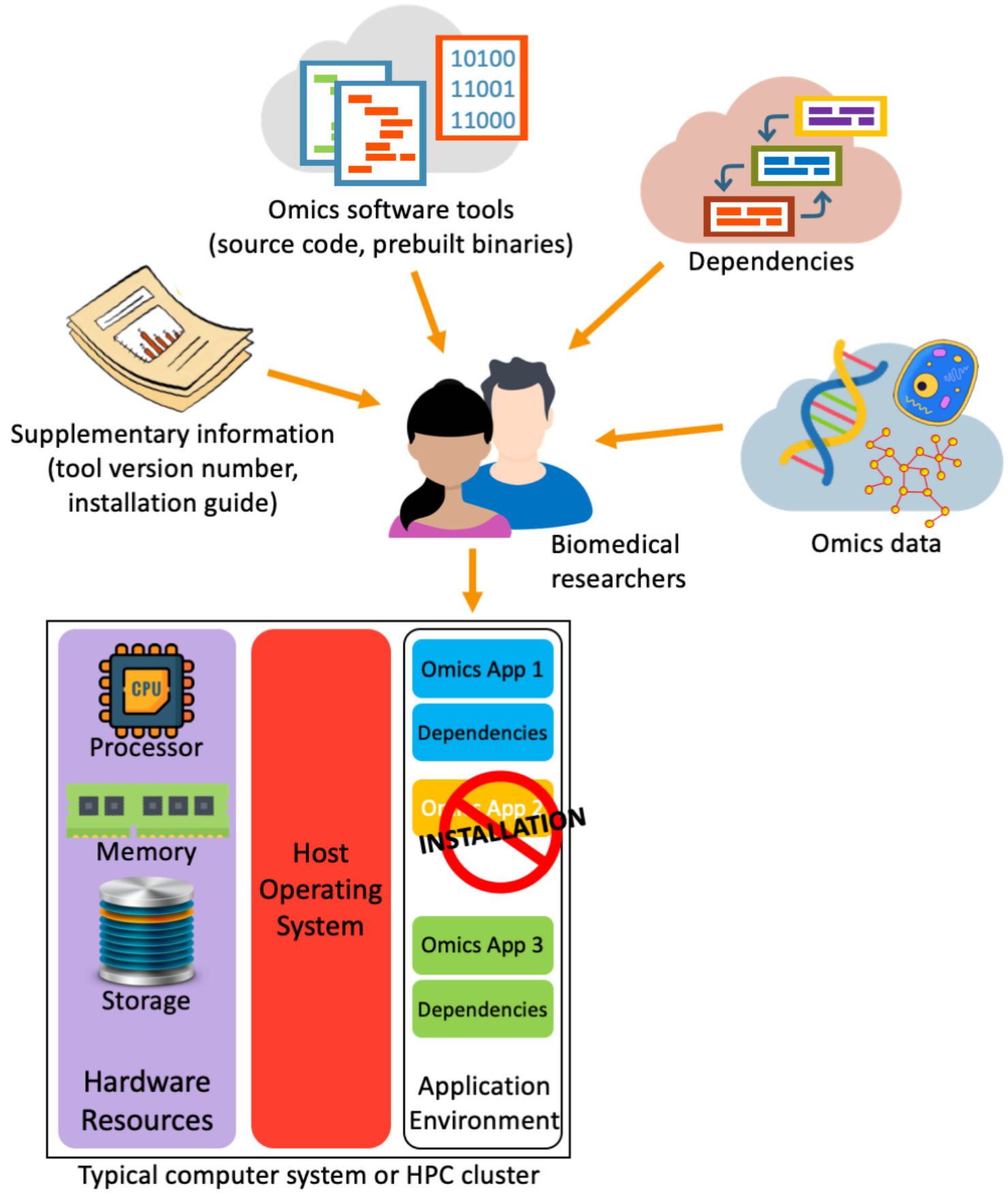

**Figure 1: General overview of the installation process for omics software tools.** When biomedical researchers need to use omics software tools and reproduce reported results, they first need to locate the version number and web address of each omics tool using the published research paper and supplementary information. They can then download the omics tools, determine each tool's dependencies using information provided by the tool's developers, and try to install the tools on their personal computer or HPC cluster. If the tool is successfully installed, the researchers download the relevant omics data and apply the tools as needed. However, even when the hardware resource requirements (CPU type, memory capacity, and storage) of each tool are met, some omics tools are likely to fail when exact reproduction is attempted because of installation challenges.

These difficulties reflect an understanding gap between users and software developers, which is directly observable in the lack of a user-friendly interface for many academic software tools[2,21]. Another common problem is that different software tools may require different versions of the same dependency, which can cause conflicts with existing software on systems that allow for only one version of a particular dependency to be installed at a time[13,20]. Furthermore, software versions often change over time, sometimes leading to differences in performance, memory footprint, and end results. If an analysis depends on recently introduced features of a tool, attempts to reproduce the analysis using an older version of the tool will fail. Likewise, programming languages can also change over time. For example, Python developers introduced a new version of the language, Python 3, that was not backward compatible with the previous version of the language, Python 2[22]. Some functions in Python 3 perform different tasks or have different configurations than their counterparts in Python 2, although they have the exact same name in both versions. Because omics tools and their dependencies use programming languages such as Python, it is necessary to keep track of the exact versions of all dependencies and programming languages used. The challenges worsen when users try to install a set of software tools in which each tool must satisfy requirements of other tools in the set, which can lead to software incompatibility[23].

Biomedical researchers who use omics tools often have access to high-performance computing (HPC) infrastructure (e.g., HPC clusters or cloud computing) but lack the skills required to install the tools on the HPC platform. An HPC cluster is a group of machines that work in tandem to provide computational resources beyond those of individual machines. HPC clusters are typically available at major universities or national research labs. Users of HPC clusters and cloud-based platforms are normally prevented from performing system-level operations such as software installation to protect the system against rapid and/or unwanted changes. This limitation makes personal computers appealing for users of some omics software tools, as they can install software on computers they own themselves. However, users of both HPC clusters and personal computers still encounter the other common challenges that we discussed above when installing omics software. It is crucial to address the challenges related to omics software installation on HPC platforms, as well as on personal computers, so that biomedical scientists can spend time on research, rather than installing and troubleshooting computational tools.

One approach to addressing these challenges is to wrap omics software tools in an isolated software environment using packaging[24,25], virtualization[26–28], or containerization[29,30]. Each of these approaches helps simplify omics tool deployment, while increasing the accessibility and reproducibility of biomedical research[29,31]. Packaging automates the process of downloading, installing, and configuring software tools and dependencies. Virtualization enables sharing of pre-installed software tools together with a complete application environment and OS across multiple machines. Containerization allows software tools and their corresponding dependencies, system tools, system libraries, and settings to be distributed and used in a platform-independent manner. Although packaging, virtualization, and containerization have many similar goals, they differ in important ways that affect their appeal to various types of users (**Figure 2**).

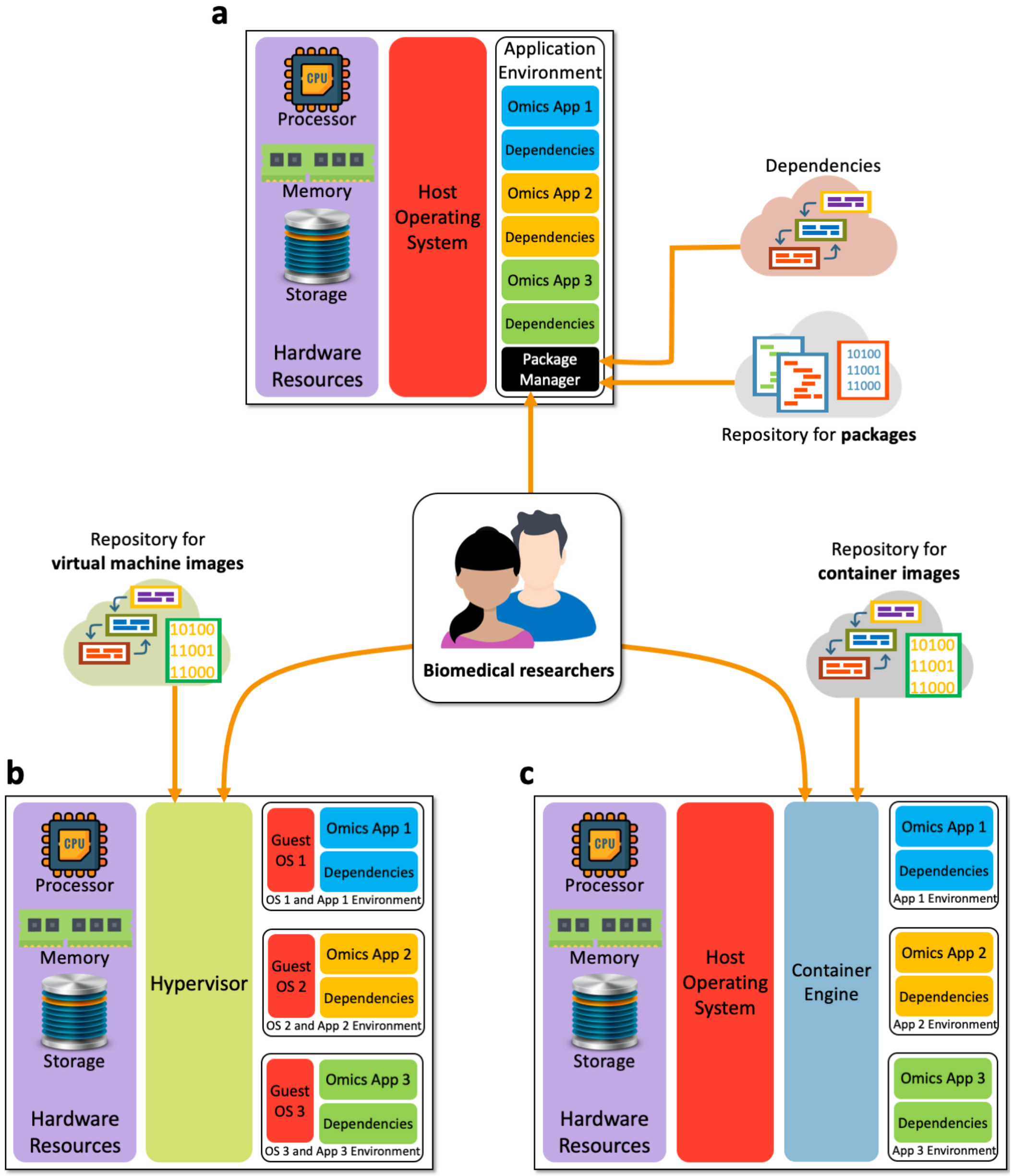

**Figure 2: An overview of packaging, virtualization, and containerization platforms for addressing challenges of omics software installation.** (a) A package includes the omics software and relevant metadata (e.g., list of dependencies and versions to be installed with the omics software) necessary to complete the installation of the packaged software. The dependencies are not included in the package and are usually obtained from sources other than the package repository. Software packaging does not provide an isolated environment for the packaged omics software tools and their dependencies, which prevents packaging from solving issues known colloquially as "dependency hell". (b) A virtual machine image includes an already-installed omics software tool along with its dependencies and operating system (OS). Virtualization provides complete isolation of the omics software from the host

OS and application environments. (c) A container image includes all binary files for the omics software tool and their dependencies along with metadata describing the OS and architecture the image was built for and the requirements needed to run the image. Containerization provides isolation from only the application environment.

There are many available platforms for packaging, virtualization, and containerization, each with different limitations and advantages. However, there is a lack of knowledge in the biomedical community about how each approach addresses specific needs for improved installability, usability, and reproducibility. Biomedical researchers come from a wide range of disciplines but often lack formal training in computer science[32,33]. In this review, we provide a comprehensive overview of existing packaging, virtualization, and containerization platforms and offer an in-depth discussion of how these platforms can be adopted to simplify the use of computational omics tools. We describe the challenges, advantages, and limitations of available packaging, virtualization, and containerization platforms from the perspectives of users, software developers, and system administrators. Additionally, we define several widely used technical terms in a **Glossary**. Our discussion has three main objectives: 1) provide information that helps biomedical researchers choose the most appropriate packaging, virtualization, or container platform for their project; 2) promote broad adoption of omics tools across a wide range of biomedical applications by providing best practices for making the tools easy to install and use; and 3) improve transparency, reproducibility, and rigor in computational biomedical analyses[16]. We discuss specific features of various platforms, such as ease of use, configuration flexibility, performance overhead, security issues, and compatibility with personal computers and HPC platforms. We also propose principles to make packaging and containerization of omics software more sustainable and reproducible to ultimately increase their adoption by the broad biomedical community.

**Package managers facilitate ease of use of omics software tools**

Package managers first appeared nearly 28 years ago, as software developers sought to streamline the software installation process[34,35] (**Figure 3a**). To install an omics software tool using a package manager, the user need only specify the omics software to be installed, and the package manager handles the download, installation, configuration, and dependencies. Various package managers are available for different OSs and purposes (**Table 1**). Many OSs have built-in package managers (e.g., APT in Debian Linux[36]), which may not be available on HPC clusters. Other package managers (e.g., Zero Install[37]) need to be downloaded and installed by the user. Some package managers such as Conda[38] are programming language agnostic (i.e., compatible with many programming languages), whereas others such as pip[39] for Python are designed for a particular programming language. The pip[39] package manager has the largest number of packaged software tools among the package managers discussed in this review (**Table 1**). Most package managers use their own package structure, referred to as the format, based on the target OS and the availability of the packaged software's source code[40]. Packages can be converted from one format to another if needed using converter tools (e.g., Alien[41]). Omics packages normally contain the omics software to be installed, a list of the necessary dependencies, and metadata such as descriptions of the packaged software (**Figure 1a**). Some omics packages, such as those on Bioconda, are free-of-charge and publicly available, whereas others have limited availability, as they are commercialized or unlicensed[42,43].

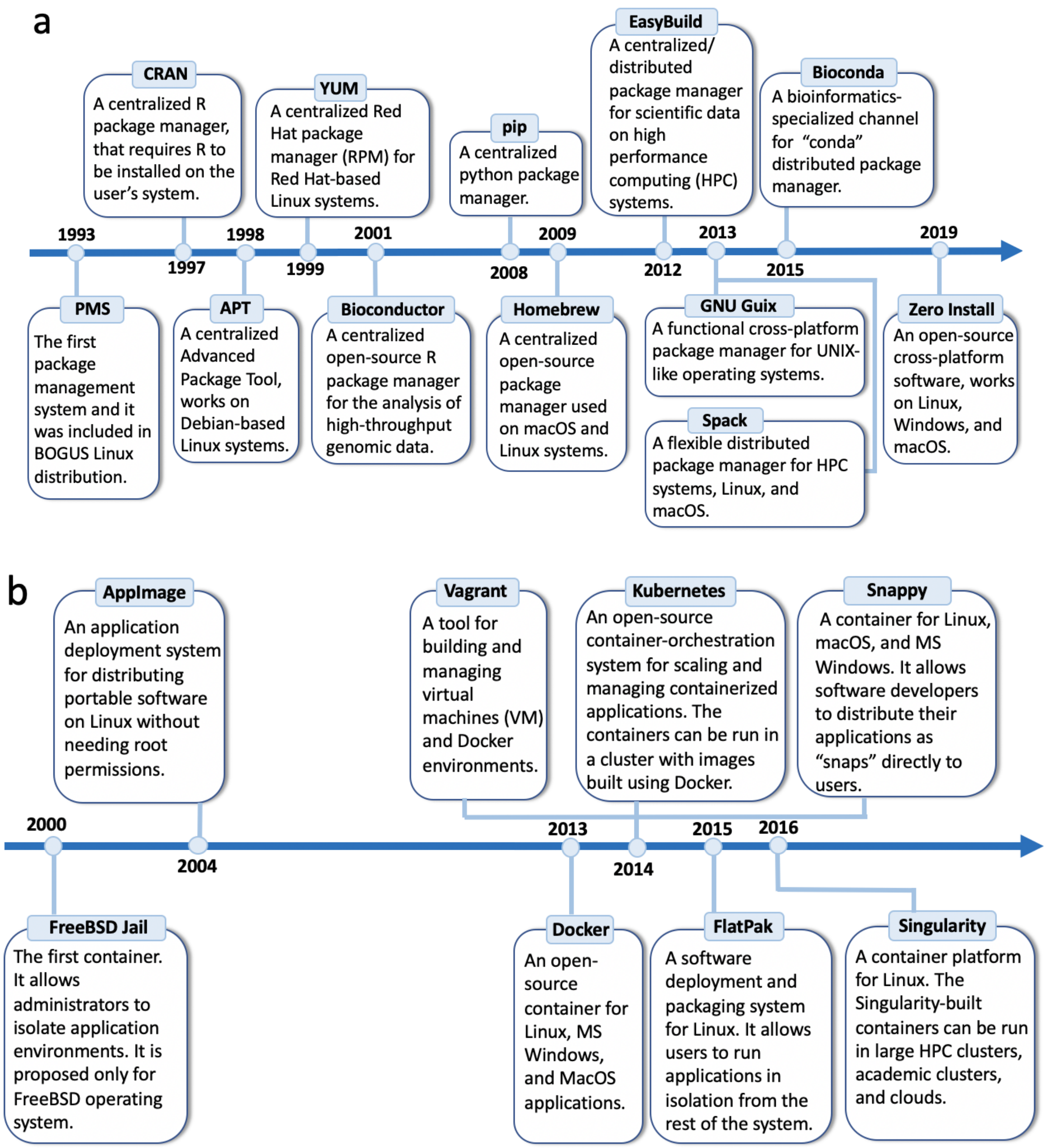

**Figure 3: Development timeline and brief description of popular (a) package managers and (b) containers.** Each tool is described with key information regarding its functionality, purpose, and supported operating system. In addition to the surveyed package managers and containers, the first package manager, PMS, and the first container, FreeBSD Jail, are shown.

To use a package manager, a user simply instructs the package manager which omics software to install, and the package manager performs all the necessary steps to install the software (**Supplementary Figure 1**). Assuming the desired software is not already installed, the package manager retrieves the appropriate package from a repository. A repository, or registry, is a server that stores all the packages available to the package manager. The package manager then uses the metadata in the package as instructions for completing the installation. For most packages, these tasks include verifying that all the required dependencies are installed. In most cases, the package manager can install any missing dependencies, which are treated as additional packages that the package manager downloads and installs. There are, however, two common problems that a package manager cannot address. The first is known as circular dependency[44], which occurs when one package depends on another package that depends on the first package. In this case, the package manager cannot install either package without first installing the other package. The second problem is known as version conflict[45], which occurs when two or more packages use different versions of the same dependency. Many package managers allow only one version of a dependency to be installed at a time, thus preventing multiple packages with version conflicts from having their dependencies met simultaneously.

**Table 1: Summary of features of popular package managers.** The table is sorted by year of initial release. We collected any major publication of theirs ("Name"), their websites ("Website"), and initial and recent release years ("Initial Release" and "Recent Release"). We also found the number of packages offered as of 1 January 2022, which is subject to increase as more packages are introduced ("Number of Tools"). We define Centralized package managers as readily downloadable binary files, whereas Distributed package managers that must be assembled but are more customizable than pre-assembled binary files ("Type"). Each package manager was also identified as being cloud compatible or not ("Cloud Compatible"), meaning it can operate in a cloud environment. We also looked at whether the package manager requires sudo permissions ("Administrator Permissions"). Lastly, we looked at what other requirements were needed for each package manager before it can be used ("Licensing" and "System Requirements").

| Name | Website | Initial Release | Recent Release | Number of Tools | Type | Cloud Compatible | Admin. Req'd? | Licensing | System Requirements |
|---|---|---|---|---|---|---|---|---|---|
| CRAN[46] | https://cran.r-project.org/index.html | 1997 | 2021-11-1 | 18,660 | Centralized | Yes | No | GNU GPL | CRAN is exclusively an R package manager, it requires R to be installed on the user's system. |
| APT | https://wiki. | 1998 | 2021- | 60,000 | Centralize | N/A | Yes | GNU GPL | For Debian based Linux |
| | debian.org/Apt | | 11-17 | | d | | | 2+ | systems |
| YUM | http://yum.baseurl.org/ | 1999 | 2011-11-28 | | Centralized | Yes | Yes | GNU GPL 2 | For Red Hat based Linux systems |
| Bioconductor[47] | https://www.bioconductor.org/ | 2001 | 2021-10-27 | 2,083 | Centralized | Yes | No | Open source | Bioconductor is exclusively an R package manager for genomic analysis |
| pip | https://pypi.org/project/pip/ | 2008 | 2021-10-22 | 348,149 | Centralized | Yes | No | MIT License | pip is exclusively a python package manager |
| Homebrew | https://docs.brew.sh/Homebrew-on-Linux | 2009 | 2022-01-04 | 5,916 | Centralized | Yes | No | BSD 2-Clause Simplified License | Used mainly for macOS systems, but also works for Linux systems |
| EasyBuild[48] | https://easybuilders.github.io/easybuild/ | 2012 | 2021-12-13 | 2,575 | Both | Yes | No | GNU GPL 2 | HPC environment |
| Conda | https://conda.io | 2012 | 2021-11-22 | 2,587 | Centralized | Yes | No | 3-clause BSD license | For Windows, macOS and Linux, and for any programming languages. |
| GNU Guix[49] | https://www.gnu.org/software/guix/ | 2013 | 2021-05-11 | 19,729 | Distributed | Yes | Yes | GNU AGPL | HPC environment |
| Spack[50] | https://spack.io | 2013 | 2021-12-23 | 6,139 | Distributed | Yes | No | MIT License or Apache License | HPC environment |
| Zero Install[37] | https://0install.net | 2019 | 2020-05-4 | 2,212 | Distributed | Yes | No | GNU LGPL 2.1+ | Libraries can be shared between programs |

**Omics software tools are largely available through centralized package managers**

Package managers are centralized if they rely on a package distributor, which collects packages from developers, approves them, and releases them in a repository (**Table 1**). An example of a commonly used centralized omics package repository is Bioconda[51]. Bioconda only contains omics software tools and requires installation of the Conda[38] package manager to use. The use of centralized package managers such as Bioconda makes omics tools easily searchable and accessible to users. As an alternative to centralized package managers, distributed package managers provide an interface for users to build their own packages from source code in a user-centric way. Examples of this approach are Spack[50] and EasyBuild[48], which are normally used on HPC clusters[52]. Centralized package managers provide two key benefits compared with distributed package managers. First, they are easier to use, as their packages are widely available as ready-to-use downloadable files. Second, they help maintain global consistency of their packages by imposing guidelines for the hosting of new packages in their repository[53]. On the other hand, the advantage of distributed package managers is that users can build their own packages that optimize performance on a given system (i.e., by using different build options, computing architectures, or compilers) without the censorship of a repository[54]. However, this optimization requires programming and other computational skills, which makes distributed package managers unsuitable for users with limited computational experience.

**Advantages and limitations of package managers for omics computational tools**

Advantages

- Package managers enable users to easily search for packages within a common repository (e.g., Bioconda), eliminating the inconvenience of searching for packages in different locations.
- Most package managers do not require administrative privileges to install their software.
- Most packages are small in size, as they include only the omics software and a list of dependencies.

Limitations

- Package managers cannot address circular dependency and dependency version conflict.
- Package managers depend on public network access search for packages in a repository. Although this can be mitigated by separately downloading the package from the repository and moving it to a machine with no internet access, it may be challenging for inexperienced users, especially when dependency issues are present.
- A package manager is not needed if there are no complex dependencies.

**Virtualization enables complete portability across different machines and platforms**

Even with a package manager, portability can be a problem. Portability refers to the ability of a piece of software to function the same on different types of machines or OSs. Virtualization enables portability by maintaining consistent software functionality, regardless of the machine on which the software is run. Virtualization first appeared nearly 30 years before the first package managers[26,27,55]. The OS of the host machine usually manages and allocates computational resources (central processing unit [CPU], peripherals, memory, and storage) to each executed software. In virtualization, abstraction of computational resources allows the OS of the host machine to share these resources with multiple guest OSs, called virtual machines (VMs), each acting as an independent computer with its own OS (**Figure 3b**). The abstraction techniques partition the computational resources and place a control software, called hypervisor[56], between the OS and the underlying hardware. Each VM operates under the illusion of exclusive access to its own resources with the help of the hypervisor. For example, a computer running a Linux OS may host multiple VMs concurrently, each running a different version of Microsoft Windows OS while having its own resources.

Virtualization provides many benefits for biomedical research. First, the host machine treats each VM as a single file, called an "image", that can be stored, retrieved, and shared. Second, developers can build images that contain pre-installed software, along with an appropriate OS and all necessary dependencies. Users without advanced computational skills can directly run the image on any machine, while maintaining the original environment and functionality of the pre-installed software. Third, users can avoid the difficulties associated with software installation and the administration of multiple physical machines with different OSs. Fourth, virtualization increases the efficiency of resource utilization by allowing multiple VMs to share the same underlying resources concurrently.

The benefits of virtualization come with computational costs. Traditional VMs, known as 'type-1 hypervisors'[56], emulate an OS and have direct access to the host machine's hardware (**Figure 1b**). The most commonly used type-1 hypervisors are kernel virtual machine (KVM), Xen Server, VMware Server, and Hyper-V. The more recently developed 'type-2 hypervisors'[56] run as a software program within the host OS by translating all host OS instructions to those that are understood by the guest image. The most commonly used type-2 hypervisors are VMware Workstation Player, VirtualBox, and Parallels Desktop. Although both types of hypervisors excel in portability, they use significant CPU and memory resources[57], which prompted the development of a newer approach called OS-level virtualization, also known as containerization. Containers allow multiple isolated user spaces to be run in parallel while incurring lower CPU, memory, and networking overhead than VMs[57].

**Advantages and limitations of virtualization for computational omics tools**

Advantages

- Virtualization enables complete portability of the omics tool, along with the entire OS that the tool needs to run.
- Portability does not require a specific host OS, meaning that the host OS and VM OS can be different.

Limitations

- Virtualization requires large amounts of CPU and memory resources because it abstracts the entire OS. This often causes the VM to consume all of its allocated resources, so it is usually not optimal to run several VMs on the same host.
- VM image sizes may be large enough to preclude easy sharing across networks.

**Containerization has lower computational costs than virtualization, while maintaining portability and ease of use**

Containerization was first introduced in the FreeBSD OS in 2000[58] (**Figure 3b**) to facilitate portability and ease of use of computational tools. Containerization allows omics software tools to be run in isolated user spaces called containers, using the same OS as the host system. The user downloads a container "image" that includes the omics software, its dependencies, and anything else necessary to run the omics software (**Figure 2c**). Although container images are not typically installed on the host machine (as it does not include an OS, unlike a VM), many require the installation of a container "engine" (sometimes called a "runtime"). The container engine is an application that manages the components required to run container images by creating an isolated application environment for each image on top of the host OS. The container engine then runs the imaged software inside the application environment[59] (**Supplementary Figure 2**). Thus, a user can install a single container engine and use it to launch numerous container images, each with a different omics software tool. This design makes the images both highly portable and easily shareable across standard network connections, which has led to wide adoption of containerization by users and developers of omics tools[60,61].

Containers solve two fundamental problems that users of omics software face. First, they package all the dependencies into a virtual environment where the omics software can function without requiring the user to download or configure any additional components. Second, they avoid potential problems due to version incompatibilities by running each omics tool in its own container with its own separate set of dependencies. Each container is independent of other containers and has its own independent file system, meaning there can be no overlap or dependency errors.

Docker[62] is the most popular containerization platform and also offers the largest number of container images (both omics and non-omics tools) for download (**Table 2**). Developers can build images by creating a plain-text build recipe (called Dockerfile) that specifies whether to build on a previous Docker image (referred to as parent image) or to start from new contents (referred to as base image) and which software libraries and dependencies should be installed. The Dockerfile also provides easy-to-read documentation of the dependencies that are required[29]. Docker then builds an image based on the developer's specifications that can be uploaded to the Docker repository[63]. If the image is released publicly, users can download the image and execute it as a container on their own OS. Different versions of the same software can be built, released, and downloaded as necessary. This approach makes it easier for developers to build and deploy software that works the same on any Linux machine. Users benefit by being able to quickly download pre-built images with confidence that they are receiving the tool they want, packaged with the exact dependencies specified by the developer. Containers usually incur higher security concerns than package managers and virtualization methods. Container images that do not follow up-to-date best practices are vulnerable to malicious attacks[64–66]requires giving users high-level system access, which can be exploited to compromise the system.

**Table 2: Summary of the features of popular container platforms.** The table is sorted by year of initial release. We collected any major publication of theirs ("Name"), their websites ("Website"), and initial release years ("Year"). We also found the number of container images offered as of 1 January 2022, which is subject to increase as more images are introduced ("Number of Tools"). Each container was also identified as being cloud compatible or not ("Cloud Compatible"), meaning that the platform can operate in a cloud environment. We also looked at whether the container requires sudo permissions ("Administrator Permissions"). Lastly, we looked at what other requirements were needed for each container before it can be used ("System Requirements").

| Name | Website | Year | Number of tools available | Cloud compatible | Administrator Permissions | Notable Features |
|---|---|---|---|---|---|---|
| *AppImage* | https://appimage.org | 2004 | 1,303 | Yes | No | Linux-distribution agnostic |
| *Docker*[62] | https://www.docker.com | 2013 | 8,691,452 | Yes | Yes | Operating system agnostic |
| *Vagrant*[67] | https://www.vagrantup.com | 2014 | N/A | Yes | No | In 2010, Vagrant was used for managing VM environments. In 2014, it started managing docker environments. |
| *Snappy* | https://snapcraft.io | 2014 | 4,725 | No | No | Operating system agnostic |
| *Kubernetes (K8s)* | https://kubernetes.io | 2014 | N/A | Yes | No | Kubernetes helps manage Docker containers in an HPC cluster |
| *FlatPak* | https://flatpak.org | 2015 | 1,366 | No | No | Linux-distribution agnostic |
| *Singularity*[68] | https://sylabs.io | 2016 | 100 | Yes | No | Designed for Linux and HPC systems. It can convert Docker images to Singularity image format. |

**Advantages and limitations of containerization for computational omics tools**

Advantages:

- Container images provide a hassle-free environment for testing, deploying, and running omics tools.
- Container images include all dependencies in a single image to ensure that the omics tool runs on the host OS and architecture.
- Developers can easily deploy images for different types of OSs and architectures.
- Containers are optimized to use minimal resources, which allows multiple containers to be run in parallel without incurring high overheads in CPU and memory.

Limitations:

- Containers have more security concerns than package managers and virtualization. Images that do not follow up-to-date best practices are vulnerable to malicious attacks[64–66], and inexperienced users may struggle to filter such vulnerable images to protect their system.

### Benefits and limitations of packaging, virtualization, and containerization to users of omics software

Although package managers, VMs, and containers seek to accomplish many of the same goals, they take different approaches that can affect their appeal to various types of users. Usability is defined by how easily a software tool can be installed and run by an everyday user, which is determined by factors such as whether users need administrator privileges and/or advanced technical skills to install and run a piece of software. Some package managers, such as Conda, pip, CRAN, and Homebrew, enable users to install software packages with a single command, whereas others, such as EasyBuild and Spack, require the user to build executable applications from source code. Package managers of the second type are typically used only by system administrators of HPC clusters. Some package managers, such as APT and YUM, require administrator privileges to use and are therefore completely unsuitable for everyday users (**Table 1**). Generally, virtualization and containerization are easier to use than package managers once the container engine or VM software is installed, and all the user needs to do is download and run images.

Package managers have a compact design, with individual packages containing only the software of interest and instructions specifying its dependencies. This makes individual packages small enough that they can be easily shared and downloaded from repositories. If a dependency is missing, the user must download and install it separately. Under most circumstances, this allows for only one version of a particular piece of software or dependency to be installed at a given time, but the installed version is available to any other software that requires it. Package managers fail to address problems such as circular dependencies[44] and conflicting versions[45], known colloquially as "dependency hell"[69]. By contrast, containerization and virtualization are based on a fully integrated design with all dependencies included in a single image, together with the software of interest, which means that the correct version of each dependency is always available once the image is installed.

The usability of packages, VMs, and containers is limited if users cannot easily locate and access the software they need. Package managers make it easy to search for packages online, and there are omics-specific repositories such as BioBuilds[70] (**Table 3**). Containers are not searchable on the internet in the way packages are, so they are usually stored in searchable repositories such as the domain-agnostic Docker Hub[63] and omics-specific repositories such as Dockstore[71] and Bioboxes[72]. There are also omics-specific repositories for VM images, such as Bio-Linux[28]. Some repositories, such as BioContainers[60], include both packages and containers to further facilitate the searchability and accessibility of omics software. Repositories of packages and containers do not actually perform the packaging or containerization but rather serve as a library for users to download packages and container images that are created by developers and then stored in the repository.

**Table 3: Summary of available repositories that house only omics software tools.** These repositories do not perform the actual packaging or containerization. They serve as a library for users to download already created packages and container images. We collected their major publications ("Name"), websites ("Website"), and initial release years ("Year"). We define the type of omics tools that are included in each repository ("Type of Tools") and the number of software tools in each repository as of 1 January 2022 ("Number of Tools").

| Name | Website | Year | Type of Tools | Number of Tools |
|---|---|---|---|---|
| Bioconductor[47] | https://www.bioconductor.org/packages | 2001 | Packages | 2,083 |
| Bio-Linux[28] | http://nebc.nerc.ac.uk/nebc_website_frozen/nebc.nerc.ac.uk/tools/bio-linux-7 | 2006 | VM images | 500 |
| Bioconda[51] | https://bioconda.github.io | 2015 | Packages | 9,057 |
| BioContainers[60] | https://biocontainers.pro | 2015 | Packages and container images | 215,300 |
| Dockstore[71] | https://dockstore.org/search | 2015 | Container images | 200 |
| Bioboxes[72] | http://bioboxes.org | 2015 | Container images | 9 |
| BioBuilds | https://www.l7informatics.com/resources/biobuilds-home/ | 2016 | Packages | 129 |

Because omics datasets can be very large and require substantial time and computational resources to analyze, the usability of omics tools may depend on the performance of the tools on the host machine. The distributed package managers Spack and EasyBuild provide a relatively high level of performance because their settings can be optimized to fit the host system, and their packages are built from source code to fit the host hardware. These distributed package managers also allow multiple versions of the same software to be installed at once. Although the centralized package manager Conda cannot be optimized to the extent that Spack and EasyBuild can, it can still take advantage of high-performance libraries such as the Intel Math Kernel Library (MKL). Other centralized package managers, such as APT, YUM, Homebrew, pip, and CRAN, that contain only prebuilt binary files have the lowest performance, as the prebuilt files are not usually optimized for the machines that will run them. Virtualization and containerization both introduce performance overheads that, especially in the case of virtualization, can limit their usability in performance-critical tasks. Because containerization has direct access to the OS kernel, its performance overhead is substantially less than that of virtualization and comparable to that of non-containerized software[57,73–75].

User privileges can affect the usability of package managers (centralized and distributed) and containers on HPC clusters that share resources among many users. Most users' access to HPC clusters is limited to protect the system from security issues. Therefore, when a user wants to use an omics tool on an HPC cluster, the user may need to ask a system administrator to install the needed package or container. The administrator might deny the request for various reasons, such as policy conflicts, software conflicts, permission conflicts, or security risks[76]. Furthermore, the use of containers on HPC clusters generally requires a safe container format, such as the Singularity runtime. For example, Docker generally is not allowed to be used on HPC clusters because it requires giving users high-level system access; however, this can be resolved by using Singularity to launch the images.

**Benefits and limitations of packaging, virtualization, and containerization to software developers**

For software developers, package management and containerization are closely aligned. It is crucial that software tools be provided through an easy-to-use, unified interface so that they can be easily deployed and managed by users. Omics tools that are not available as packages or container images can be extremely challenging to install[13,20]. Therefore, developers build packages and containers using centralized, versioned, and shareable software so that users do not have to build applications from source code or find applications that have binary compatibility with the host computer. This alleviates the need for developers to provide version-specific source code, detailed build instructions (taking into account the OS version and distribution), and any required dependencies. In this way, developers can spend more of their time developing software relevant to their field of study and less time on peripheral activities. Furthermore, containerized software is more likely to be adopted by users than software that is difficult to install and run because containerization saves users' time and effort and allows them to concentrate on their research activities. The benefits of easy deployment and management come at cost, however, because developers, who can be relatively new to software development, must acquire specific computational skills to package or containerize their software.

Package managers typically act on the entire host OS, so when one user installs a package, the package is installed for all users of that computer. This lack of isolation can create difficulties for developers. When developers create different software tools for different contexts, they commonly rely on entirely different sets of dependencies, which can be version-specific and may clash with other dependencies that were installed for other purposes. Some package managers try to address this directly (e.g., the Python languages concept of the VirtualEnv), but not all do. Another limitation of package managers is that developers depend on the curated package repositories built into various OS distributions. To get unsupported packages or versions, developers need to use custom or unsupported repositories, which can threaten system stability. The underlying problem with regards to both of these limitations is the absence of an isolated application environment for the software of interest (i.e., the omics tool), which can be overcome by using containerization instead of package management.

The standard deployment process for containers makes containerized software more portable than non-containerized software (**Supplementary Figure 2**). Omics tools that are created and tested on a researcher's personal computer and then containerized can be reliably deployed in an institutional setting, or in the cloud, because the container provides abstraction from the underlying infrastructure and OS. The use of third-party tools and libraries in software development also becomes simplified with containerization. When software tools are made available as container images, they no longer need to be installed on a host device in order to run on that device, nor do they need to be compiled separately for multiple platforms. This bypasses issues of clashing dependency versions or complex installation steps.

Containers can be considered a type of lightweight virtualization, providing developers with the same kind of stable, reproducible environment that virtualization provides but with the additional and vital advantage of being inexpensively and easily shared across networks. Containers can be deployed locally or on the cloud using standard procedures and standard platforms, such as Kubernetes, Amazon's Elastic Container Service, or the serverless AWS Lambda platform. This enables containers to be used as collaborative spaces[77]. Specifically, the recipe that creates a container image (e.g., a Dockerfile) can be shared among developers with different skill sets (i.e., a bioinformatician and a software engineer). In this way, each developer can contribute their specific knowledge to build containers that support and run complex omics tools with multiple components, such as bioinformatics pipelines. Containerization thus enhances the ability of developers to collaborate across domains and abstraction layers. This is especially important for omics software development, where programs are usually written at least in part by developers whose primary field of expertise is in the life sciences, rather than in software engineering. Omics software also benefits from the contributions of software engineers, who can prepare reliable base images on top of which life-science researchers can construct programs to accomplish specialized tasks. The resulting containers can then be run relatively easily by software engineers (i.e., "devops" staff) in complex, orchestrated contexts as part of larger systems.

**Benefits of packaging, virtualization, and containerization to HPC system administrators**

For HPC system administrators, it is more efficient to build software using source code than to run prebuilt binary files because building from source code maximizes performance for the particular system on which the software will be used. HPC system administrators typically curate multiple software stacks and package versions based on different compilers and/or compiler versions (e.g., GCC versus Intel compiler) to provide a diverse array of software to users and to optimize performance. A software stack is a collection of independent software tools and libraries that work in tandem to support the execution of another software tool. On HPC clusters, a primary limitation of centralized package managers (e.g., APT or YUM) is that these package managers only support one software stack and one version of a given package. Therefore, HPC system administrators use distributed package managers such as Spack or EasyBuild that provide flexibility to build and combine various configurations. Administrators can manage the user environment and enable users to quickly switch to different configurations and manage different versions of the same software using Lmod[78]. This flexibility can improve software performance by allowing the software to better exploit the computational power of the HPC system. Although the lack of flexibility of centralized package managers is a concern for administrators of HPC clusters, it is typically not a concern and can even be an advantage for users, who do not have to manage where and how to install software tools.

Managing multiple software stacks on an HPC cluster can be time consuming, depending on the quality of the existing build recipes and the specific packages needed by users. Package managers like Spack and EasyBuild are open-source and community-supported platforms, so they already contain many build recipes provided by the community of users. However, sometimes there is no existing build recipe for an omics tool, or the existing build recipes do not work as expected on a particular HPC system, so new recipes must be developed. Though build recipes can target a specific compiler and packages, system administrators typically focus on maturity and reliability when choosing compiler and package versions to support. In addition, some HPC centers have a heterogeneous mix of computing architectures, which makes managing software for maximum performance more difficult, as administrators need to rebuild software tools from source code for each different architecture.

Containers are often the preferred solution for either legacy or cutting-edge software, as well as for software that is difficult to install. It is relatively easy for system administrators to install a container engine (e.g., Singularity), but containers place an extra burden on users to learn the basic concepts of containerization and the specifics of how to build (accounting for different image formats) and run container images. This is in contrast to software installed using a package manager, which runs like any other program on the host system once it is installed properly. For example, containerized software might be more difficult to debug than software installed with a package manager. This can place a burden on system administrators and support staff to help users build and run container images. In some cases, commonly used images may be hosted locally on an HPC system and provided to users to reduce the burden of building images, but containerized software that is maintained on HPC systems typically requires customization. Furthermore, containerized software has potential security vulnerabilities that software packages do not, so system administrators must keep them updated for the latest security fixes. For example, it can be difficult for system administrators to deploy Docker on HPC clusters because Docker containers are vulnerable to compromise, denial of service, and privilege escalation attacks[64–66]. Another challenge facing system administrators is that container engines only work with specific host OSs[79] (**Table 2**), which can limit their usability. Overall, deployment of containers on HPC clusters is a promising possibility but presents substantial engineering challenges related to resource utilization and allocation[80]. Currently, the most popular containerization solution for HPC clusters is Singularity, which converts Docker images to the Singularity image format and then runs them on the cluster.

**Packaging, virtualization, and containerization increase the usability and reproducibility of omics tools**

Modern omics software distributions are already becoming more reliant on package managers and containers. This is evident in the large number of software tools that are available from omics-specific package and container repositories[60,71,72] (**Table 3**). Two of the largest and most popular omics software repositories are Bioconda[51] and BioContainers[60], which both started in 2015. There is a similar trend in the increasing number of major omics initiatives, such as the Critical Assessment of Metagenome Interpretation challenge[6,81] (known as CAMI challenge), the National Cancer Institute’s Genomic Data Commons[82], and commercial omics service providers such as Seven Bridges Genomics[83]. Wide adoption of packaging and containerization in biomedical research will necessitate strict standards on the input and output of omics software tools. This will simplify the process of assembling containers into omics pipelines and make it possible to replace certain tools if necessary. Bioboxes has already established a platform for collection of Docker-based containers with standardized interfaces to make omics software interchangeable[72]. In order to make omics software tools easily searchable and accessible, Global Alliance for Genomics and Health[84] developed a standardized interface to examine the availability of omics tools in repositories such as BioContainers[60] and Dockstore[71]. Containers do not virtualize hardware, so they have a smaller computational footprint than VMs[85]. Although Docker images are typically small and lightweight, omics datasets are often very large[86], so inclusion of omics data inside a container or VM image dramatically increases the image size and limits its deployment and shareability. Therefore, omics data are usually not included in container or VM images but are instead shared through publicly accessible data repositories.

Package managers and containers can be integrated into existing workflow management systems to create sustainable interfaces that support computational reproducibility of biomedical research[15,87]. The reproducibility of results obtained by computational omics methods is extremely important, especially if the results are used in clinical settings. In terms of reproducibility, the relationship between packaging and containerization can be seen as that between two concentric circles. At their core, package managers simplify and standardize the installation of dependencies so that the dependencies are the same across multiple installations of the same software. Containers add a further wrapper around this by doing the same thing for the directory structures and files that the software expects to find at runtime. Moreover, considering that containers are typically defined by a build recipe file, the maintenance of these recipe files under source control (tracked changes to recipe files) ensures reproducibility over time as developers make improvements to the software.

To independently reproduce an omics software process, at least three hurdles must be cleared. First, it must be possible to retrieve the exact version of the source code that was used in the omics process that is being reproduced. Second, it must be possible to build (i.e., compile and link) from the source code to produce an executable file that exactly matches the one used in the original process. Third, it must be possible to run the executable file on the platform(s) for which it was intended. These challenges and the techniques that exist to address them grow in complexity when an omics process is distributed across multiple processors and computers. This necessitates the need for universal platforms that are truly agnostic to package format, image format, programming language, and OS.

There is an increasing need for better management of different versions of omics software tools while allowing for software updates. Containers based on existing package managers might represent the most sustainable solution, as they allow software to evolve inside the package manager, while having the low computational resource overhead and ease of use of a container[88]. Although the container image represents a static object, the tools made available by the package manager can be easily and automatically updated along with their dependencies. Automation of software updates is essential in biomedical research, as algorithm development must accommodate rapid changes in technology[1,89]. However, software updates can hinder reproducibility and lead to usability failures when an update to one software tool creates incompatibility with other software that relies on the updated tool[23]. Biomedical researchers should be able to retrieve specific versions (e.g., as reported in a research paper) of a tool, rather than always accepting the latest version. This can be achieved by building packages from source code or recipes that use a specific software version and then sharing the source code or recipes with the biomedical community instead of sharing the image itself[77]. The majority of omics tools are available through centralized, well-maintained package managers such as Bioconda[51] and Bioconductor[47]. As the maintainers of these platforms review all of the packages hosted by the platforms, the hosted packages are always installable[20]. However, this management can delay package updates and limit the freedom of developers to make changes[90]. For example, the list of omics software tools included in Bioconductor[47] is updated only twice a year. Sustainability also requires reliance on free-of-charge solutions that all biomedical researchers can use. Despite the numerous advantages of package managers and containers, the sustainability of those solutions remains an open question.

## Discussion

Packaging, virtualization, and containerization support easy installability, usability, and reproducibility of computational omics tools. By wrapping omics tools into an isolated software environment, these platforms make omics tools easy to install, agnostic to programming language and OS, and often usable on HPC clusters without administrative privilege. Developers of omics software tools are responsible for ensuring that researchers can use their tools to get reproducible results[91]. It usually takes far less time for developers to provide instructions for installing and running their tools, and in some cases reproducing their results, than for users to figure out these steps on their own. Hence, many stakeholders have an interest in ensuring that standards are met, including educators, academic publishers, scientific funding agencies, and repositories that accommodate source codes.

The current workflow of omics software development in academia and industry encourages researchers to develop and publish new tools, but there are no commonly accepted principles ensuring the ease of use of published software[13,20,92]. This results in many omics tools requiring manual interventions (e.g., installation of dependencies or ad hoc editing of computer code). Manual interventions and long installation times are unappealing to most users, especially those with limited computational skills. The long-term maintenance and improvement of omics software tool increases its impact and usability[93]. Furthermore, the skills necessary for software installation are generally not part of the traditional life science curriculum at major universities. Professional development tutorials[94] on software packaging provided at international omics conferences may help to alleviate this critical skills shortage, but these tutorials typically have a limited audience because of registration costs and geographical barriers[95]. This may require securing positions for software developers in academia and encouraging further funding for software maintenance and improvement—not just developing new tools[96].

Scientific journals have introduced rigorous requirements for the sharing of data and source code; however, there are currently no effective requirements for the installability of published software tools and the reproducibility of results based on such tools[97]. This creates unfortunate situations in which researchers share source code that either does not run and/or produces different results when processes are repeated[20]. Thus, new standards supporting the use of packaging, virtualization, and containerization for omics software tools are needed to pave the way for biomedical research to become more transparent and verifiable. We believe such standards will increase the continuity and scientific impact of published research and enable the development of better solutions in the fast-moving arena of biomedical research.

**Acknowledgments**
We thank Michael Sarahan (Principal Software Engineer, Manager at Anaconda, Inc) for our useful discussion at the AnacondaCON 2019 conference. We thank Thiago Mosqueiro (Amazon.com) for the fruitful discussion and feedback. M.A. dedicates this paper to the memory of his father, who passed away on 9th March 2022. Our paper is also dedicated to all freedom-loving people around the world, who fight for our freedom.

**Author Contributions**
M.A. and S.M. led the project. S.M. conceived of the presented idea. M.A., R.A., N.R., S.W., N.A., and V.S. collected data. M.A., S.W., and N.A. produced the figures. M.A., S.W., R.A., B.L., C.F., R.J.A., D.S., BD.K., N.L., O.M., and S.M. wrote, reviewed, and edited the manuscript. All authors discussed the text and commented on the manuscript. All authors read and approved the final manuscript.

**Competing Interests**
R.J.A. is employed by Digital Science (digital-science.com), which develops products related to containerization and reproducible research. R.J.A. is not involved with these products, and Digital Science had no role in the study or the preparation of this manuscript. All other authors declare no competing interests.

**Funding**
O.M. and SAFARI Research Group members (M.A., C.F., N.A.) are supported by funding from Intel, VMware, Semiconductor Research Corporation, the National Institutes of Health (NIH), and the ETH Future Computing Laboratory. S.M. is supported by the National Science Foundation (NSF) grants 2041984 and 2135954.

## Glossary

- **Administrator**: A person in charge of maintaining a computer or HPC cluster.
- **Developer**: A person who writes software programs.
- **End-user**: A user for whom software is intended.
- **User:** A person, possibly an administrator, who has access to a computer.
- **Execute, Run:** To start a program on a computer.
- **Dependency**: A prerequisite of software in which a piece of software being installed requires a second piece of software to already be installed.
- **High-performance computing (HPC) cluster:** A type of supercomputer commonly found at universities and other research institutions. HPC clusters consist of many nodes, which allows them to excel at creating and processing massive amounts of data.
- **Installability:** A relative measure of how difficult it is to install a software program.
- **Interface:** A method of interacting with software.
- **Node:** An individual server that is part of an HPC cluster. Some nodes maintain the HPC cluster, while others perform complex computations.
- **Permission:** A grant or denial of access to a particular file on a computer. A user has permission if they have limited rights to a file and "full permission" if they have full rights to a file. Typically, only administrators have full permissions to every file on a computer, whereas regular users have read-only permissions to common files and full permissions only to their own files.
- **Prerequisite:** A program that must already be installed and accessible to successfully install and/or run another program.
- **Right:** An ability to take an action on a file. Common rights are 'read' (view the contents), 'write' (change the contents), and 'execute' (run the contents as a program). No right confers or precludes any other right.
- **Software:** A program or collection of programs that can be executed/run on a computer.
- **Usability**: A relative measure of how difficult it is to run a software program.
- **User-friendliness**: A measure of how easy software is to use, combining ease of installation and usability.

## Supplemental Materials

### Tables 1, 2, and 3 Methods:

**Release Dates:** Initial release dates for each of the package managers. This information was found on the websites of each package manager, or was recorded as the year that the first version was released if no official year was noted.

**Number of packages, container images, and virtual machine images:** This information was obtained from the websites of the respective platforms. If the desired number is explicitly mentioned in each website, we go through each website, download its respective HTML body that contains the page data with the list of tools. From here, using a text editor, the list of the packages for each package manager is extracted and then the length of that list is used to get the number of packages.

## Supplemental Figures

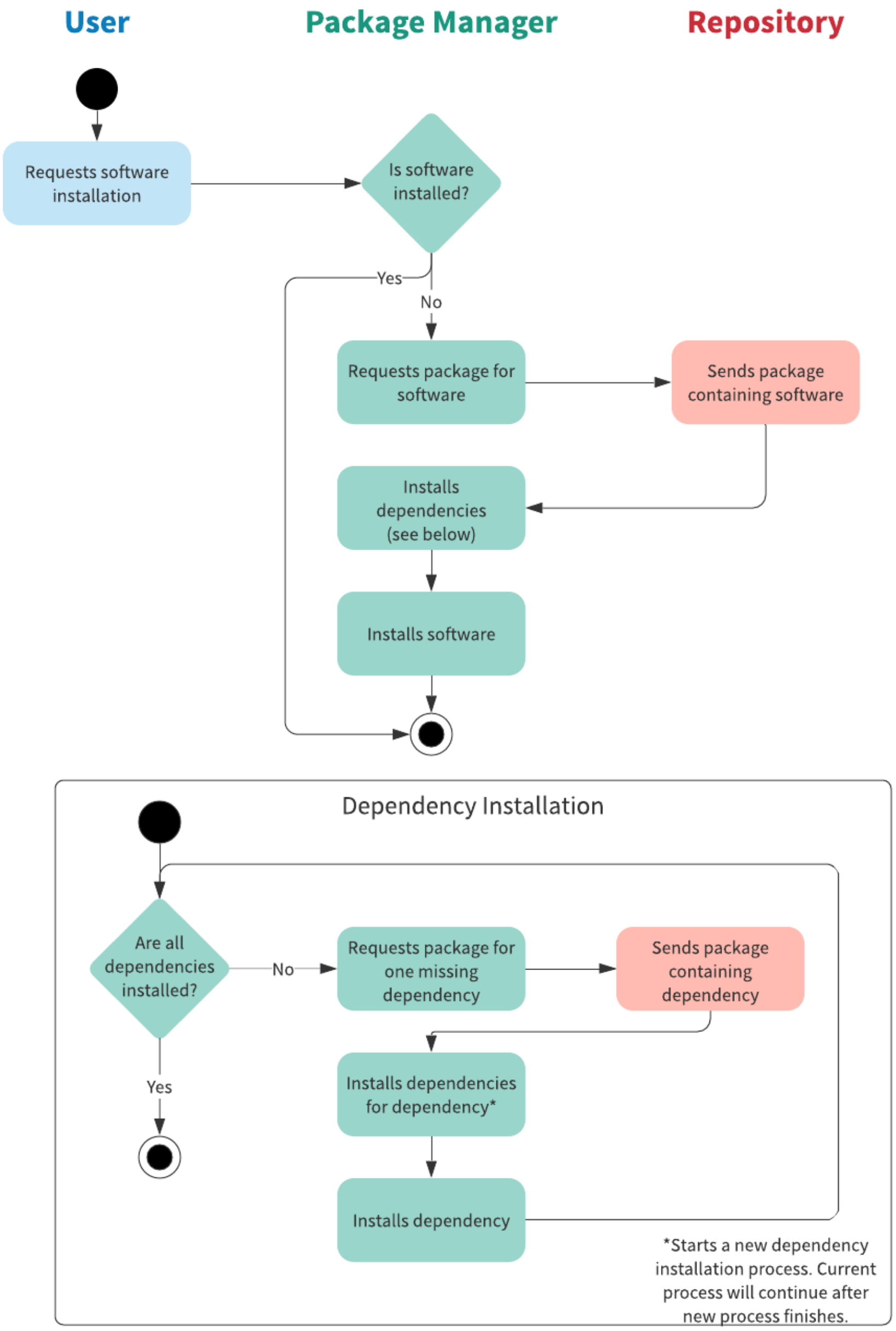

**Supplementary Figure 1: Standard workflow for installing software with a package manager.** The user, usually an administrator, asks the package manager to install a specific piece of software. If the software is not already installed, the package manager fetches the appropriate package from a repository. If any of the dependencies are not already installed, the package manager retrieves the dependency's package from the repository and starts the installation procedure for that package. Once all the dependencies are installed, the initially requested software is installed. The package manager often goes through several iterations of this process, because every dependency can have its own list of dependencies, in which case each of the dependency's dependencies must be verified through the same process.

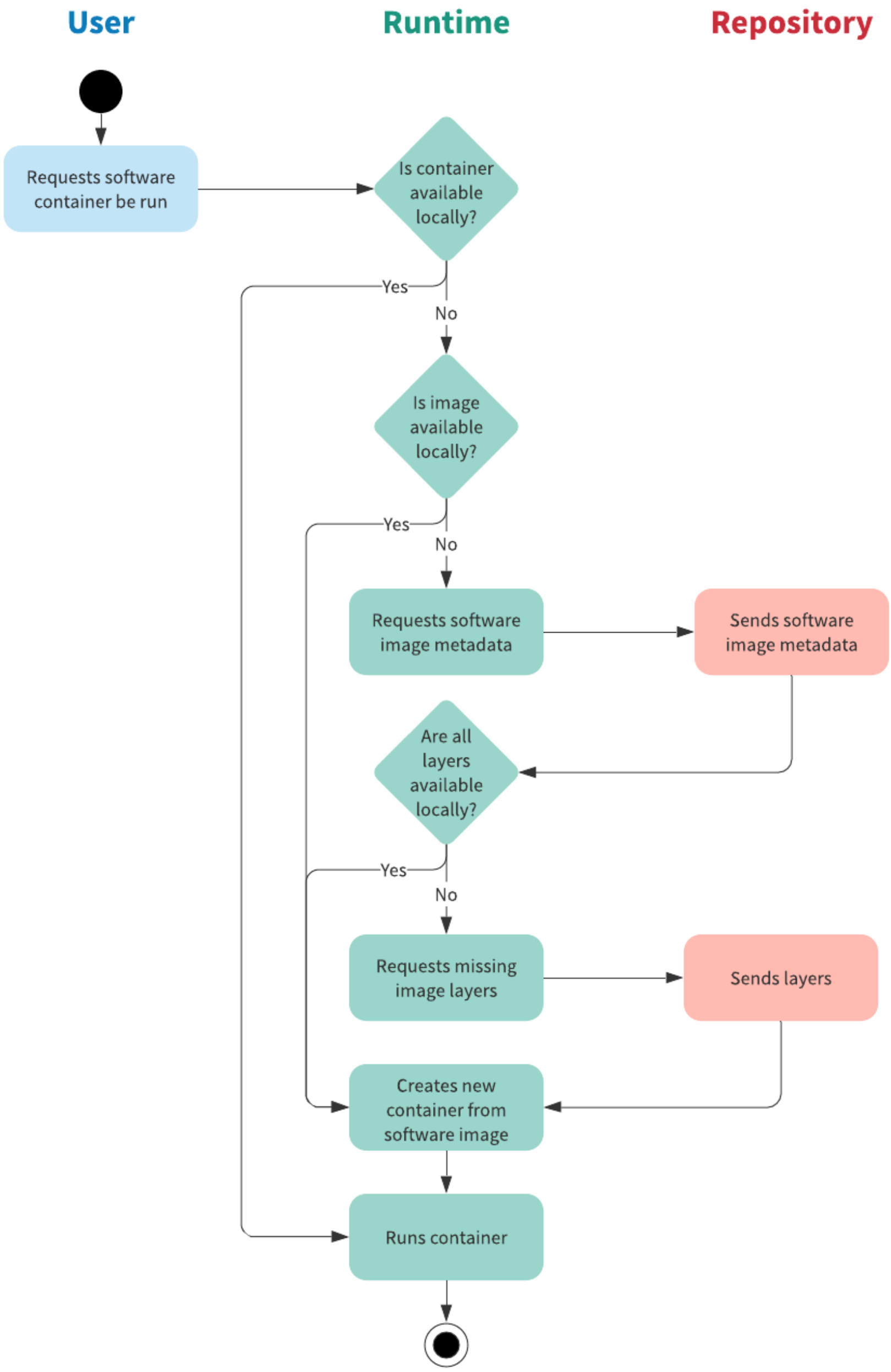

**Supplementary Figure 2: Standard workflow for running software with containerization.** The user asks to install a specific container image. If the container image is available locally, then the user can run it directly through the container engine. Potential dependencies are already handled without any intervention from the user. If the software image is not available locally, the appropriate image must be fetched from a repository.